\documentclass[12pt]{article}
\usepackage{epsf}
\usepackage{aas_macros}
\usepackage{amssymb}
\usepackage{amsmath}
\usepackage{braket}
\usepackage{mathrsfs}
\usepackage{mathtools}
\usepackage{array}
\usepackage{fancyhdr}
\usepackage{graphicx}
\usepackage{here}
\usepackage{color}
\usepackage{cite}
\usepackage{bm}
\usepackage{url}
\usepackage{fancyhdr}
\usepackage{comment}
\usepackage[colorlinks,citecolor=blue]{hyperref}
\usepackage{physics}
\usepackage{soul}
\usepackage{xcolor}
\DeclareRobustCommand{\erase}{\bgroup\markoverwith{\textcolor{red}{\rule[.5ex]{2pt}{0.4pt}}}\ULon}

\renewcommand{\thefootnote}{\fnsymbol{footnote}}
\def\thefootnote{\fnsymbol{footnote}}

\makeatletter

\@addtoreset{equation}{section}
\makeatother

\newcommand{\mrm}{\mathrm}

\newcommand{\mcal}{\mathcal}

\newcommand{\Om}{\Omega}

\newcommand{\BEA}{\begin{eqnarray}}
\newcommand{\EEA}{\end{eqnarray}}

\begin{document}

\begin{titlepage}
    \begin{center}

        \vskip .75in

        {\Large \bf Generalized early dark energy \vspace{2mm} \\ and its cosmological consequences}

        \vskip .75in

        {
            Tatsuki~Kodama$\,^1$,  Takumi~Shinohara$\,^1$,  and  Tomo~Takahashi$\,^2$
        }

        \vskip 0.25in

        \textit{
            $^{1}$Graduate School of Science and Engineering, Saga University, Saga 840-8502, Japan
            \vspace{2mm}\\
            $^{2}$Department of Physics, Saga University, Saga 840-8502, Japan
        }

    \end{center}
    \vskip .5in

    \begin{abstract}
        We investigate cosmological consequences of a generalized early dark energy (EDE) model where a scalar field behaves as dark energy at various cosmological epochs for a broad range of parameters such as the energy scale and the initial field value. We consider power-law and axion-type potentials for such an EDE field and study how it affects the cosmological evolution. We show that gravitational wave background can be significantly enhanced to be detected in future observations such as LISA and DECIGO in some parameter space.  Implications of the EDE model are also discussed for a scenario where a blue-tilted inflationary tensor power spectrum can explain the recent NANOGrav 15-year signal. We argue that the bounds on the reheating temperature can be relaxed compared to the case of the standard thermal history. 
    \end{abstract}

\end{titlepage}

\renewcommand{\thepage}{\arabic{page}}
\setcounter{page}{1}
\renewcommand{\thefootnote}{\#\arabic{footnote}}
\setcounter{footnote}{0}

\section{Introduction}

Scalar fields play an important role in various aspects of cosmology. A prime example is the inflation where a scalar field, called inflaton, drives the inflationary expansion and gives the origin of density fluctuations in the Universe\footnote{
Even if a scalar field is subdominant during inflation, such a scalar field can generate density fluctuations as in the curvaton scenario \cite{Enqvist:2001zp,Lyth:2001nq,Moroi:2001ct}, modulated reheating \cite{Dvali:2003em,Kofman:2003nx} and so on.
}. Another one is a quintessence field which can explain dark energy of the Universe\footnote{
In models with a canonical quintessence, the Hubble constant tends to be even lower than that in the $\Lambda$CDM model when fitted to cosmological data such as cosmic microwave background and so on (see, {\it e.g.}, \cite{Banerjee:2020xcn}), and hence it may not be well motivated from the viewpoint of the $H_0$ tension.
}. Scalar fields could also affect the evolution of the Universe not only during inflation and the current accelerating Universe, but also some time in between.  Indeed, high energy theories such as superstring and those with supersymmetry and so on predict the existence of scalar fields and hence they are expected to be ubiquitous in the early Universe. One of such an example is the moduli field \cite{Coughlan:1983ci,Moroi:1994rs,Kawasaki:1995cy,Moroi:1999zb} which may dominate the Universe at some epoch between the end of inflation and big bang nucleosynthesis, and could affect the cosmological evolution. 
Yet another example is an early dark energy model (see \cite{Kamionkowski:2022pkx,Poulin:2023lkg} for a recent review and the references therein, and see, {\it e.g.,} \cite{Vagnozzi:2021gjh} for possible problems in the EDE model)  where a scalar field gives some contribution to the total energy density at around the radiation-matter equality epoch, which may help to resolve the so-called Hubble tension (see, {\it e.g.}, \cite{DiValentino:2021izs,Perivolaropoulos:2021jda} for the current status of the tension)\footnote{
Early dark energy may also alleviate another tension, the so-called helium anomaly where a recent measurement of primordial abundance of helium-4 by EMPRESS \cite{Matsumoto:2022tlr} suggests a non-standard cosmological scenario, which has been discussed in \cite{Takahashi:2022cpn}.
}.  Actually, a scalar field could also help to resolve the Hubble tension in a different manner. For example, there exists a model in which the time variation of the electron mass can be generated by the dynamics of a scalar field, a dilaton \cite{Barrow:2005qf}, and such a time-varying electron mass can significantly reduce the tension \cite{Sekiguchi:2020teg}. In any case, scalar fields can play an essential role during the evolution of the Universe and have been discussed in various contexts.

A typical behavior of a scalar field is such that it slowly rolls in the early Universe and then starts to oscillate around the minimum of its potential at some point. In many scenarios, the potential of such a scalar field around the minimum is assumed to be a quadratic form (or at least the quadratic term dominates around the minimum), and hence its energy density $\rho_\chi$ dilutes as $\rho_\chi \propto a^{-3}$, which is the same scaling as that of matter. However, the potential around the minimum can be different from the quadratic one, and indeed a higher order polynomial can dominate around its minimum as in the EDE scenario\footnote{
Effects of a non-quadratic potential have also been considered in different contexts. One of such examples is the curvaton model where it has been shown that the predictions for primordial non-Gaussianities can be drastically modified from the quadratic potential case \cite{Enqvist:2005pg,Enqvist:2008gk,Huang:2008zj,Enqvist:2009zf,Enqvist:2009eq,Enqvist:2009ww,Byrnes:2010xd,Fonseca:2011aa,Byrnes:2011gh,Kobayashi:2012ba}.
}. In such a case, the energy density of the scalar field dilutes faster than that of matter, {\it i.e}., $\rho_\chi \propto a^{-q}$ with $q >3$ and especially, when $q > 4$, it dilutes faster than radiation, in which the scalar field quickly disappears and becomes irrelevant for the cosmological evolution after it starts to oscillate. Indeed such a fast-diluting scalar field (a scalar field whose energy density dilutes faster than that of matter and/or radiation) is essential in the EDE model to solve  the Hubble tension and has been rigorously investigated \cite{Kamionkowski:2022pkx,Poulin:2023lkg}.

In the context of the Hubble tension,  the initial amplitude and the parameters in the potential for an EDE field are set such that it starts to oscillate around the epoch of radiation-matter equality and its energy density should give some sizable fractional contribution to the total one, and then quickly dilutes not to affect the cosmic evolution much,  which is required to resolve the Hubble tension. However, some level of fine-tuning needs to be done to realize such a situation. From a general ground, a scalar field can dominate the Universe and start to oscillate at some epoch depending on the parameter choice and its initial value. In this spirit, we in this paper consider an EDE field in a general setting to allow various possibilities for its evolution. We refer to such an EDE field as ``generalized early dark energy" and investigate its cosmological consequences\footnote{
Another possible extension of an EDE model is to assume a  general equation of state for the initial and final EDE fluid, which has been investigated in \cite{Sharma:2023kzr}.
}.  To this end, first we identify what energy scale for the potential and the initial value for the scalar field affect which epoch in the course of the history of the Universe. Indeed, in a broad range of the parameter space, such a scalar field can dominate the Universe during its slow-rolling phase, which gives rise to a short period of inflation. After it starts to oscillate, we assume that its energy density dilutes quickly such that it becomes irrelevant to the cosmological evolution as in usual EDE models as a solution to the Hubble tension. Interestingly, in such a case,  gravitational wave (GW) spectrum can be enhanced and could be detected in the future experiments. We identify the parameter space where such an enhancement occurs. We also discuss the implications of the generalized EDE for the recent result of NANOGrav 15 year data on GW background \cite{NANOGrav:2023gor,NANOGrav:2023hvm}, particularly in models where the inflationary blue-tilted tensor power spectrum can explain the NANOGrav signal.

The organization of this paper is as follows. In the next section, we describe the setup of our scenario of a generalized early dark energy field and define several quantities that facilitate our discussion. Its cosmological evolution will also be discussed in some detailed manner. Then in Section~\ref{sec:GW}, we investigate GW spectrum in such a model, and investigate its detectability in some future experiments such as LISA and DECIGO. Implications for the NANOGrav is also discussed. In the final section, conclusions and discussion are given.

\section{Evolution of generalized early dark energy}
\label{sec:background}

In this section, first we describe the setup of our scenario and summarize the formalism to investigate cosmological consequences of a generalized EDE. Then we  discuss the evolution of the EDE field and its effects on the cosmic expansion. We also investigate possible initial values and the energy scale of the EDE field from the stochastic formalism argument.

\subsection{Setup}
\label{subsec:background}
We follow the cosmic evolution from the time just after the end of inflation to the present epoch and assume that there exist a scalar field $\chi$ (a generalized EDE field), radiation and matter components in the Universe\footnote{
Although we include a cosmological constant as the late-time dark energy component to evaluate the evolution of the scale factor up to the present epoch for completeness, it is irrelevant to our arguments below.  In our calculation, the cosmological parameters are set to the ones given by {\it Planck observation 2018}\cite{Planck:2018vyg}: $h=0.6766$ and $\Om_mh^2=0.1424$ when necessary.
}. 
The equation of motion for $\chi$ and the Friedmann equation is given by
\begin{align}
    \label{eq: equation of motion chi}
    &\ddot\chi + 3H\dot\chi + V_{,\chi}(\chi) = 0\,, \\[8pt]
    &H^2 
    = \frac{\rho_{\rm tot}}{3 M_{\rm Pl}^2} 
    = \frac{1}{3 M_{\rm Pl}^2} 
    \left( 
    \rho_r + \rho_m + \rho_{\rm DE} + \rho_\chi
    \right)
    \,, 
\end{align}
where $V(\chi)$ represents a potential of the scalar field. A dot denotes a derivative with respect to the cosmic time $t$ and $V_{,\chi}(\chi) = \dd V(\chi)/ \dd\chi$, $a(t)$ is the scale factor of the Universe, normalized to be unity at present, ${H}\equiv \dot{a}/a$ is the Hubble parameter, $M_{\rm Pl} \equiv 1/\sqrt{8\pi G} \simeq 2.436 \times 10^{18}\,{\rm GeV}$ is the reduced Planck mass. $\rho_{\rm tot}$ is the total energy density and $\rho_r$, $\rho_m$ and $\rho_{\rm DE}$ are those of radiation, matter and dark energy components, respectively. $\rho_\chi$ is the energy density of $\chi$ which is given as
\begin{align}
\rho_\chi = \frac{\dot\chi^2}{2} + V(\chi)\,.
\end{align}
In this work, we consider the following two potentials for the EDE field $\chi$:
\begin{align}
    \label{eq:power-law potential}
    V(\chi) 
    &= V_0\biggl(\frac{\chi}{M_{\rm Pl}}\biggr)^p& 
    \!\!\!\!\!\!\!\!\!\!\!\!\!\!\!
    &(\mbox{power-law})\,,\\[8pt]
    \label{eq:axion-type potential}
    V(\chi) 
    &= V_0\biggl(1+\cos\frac{\chi}{f_a}\biggr)^n&
    \!\!\!\!\!\!\!\!\!\!\!\!\!\!\!
    &(\mbox{axion-type})\,,
\end{align}
where $p$ and $n$ represent the power-law index, $V_0$ is the energy scale of the potential, and $f_a$ is the decay constant. These types of potential, particularly with $p\ge 4 $ and $n \ge 2$ are well investigated in the context of the Hubble tension \cite{Poulin:2023lkg} since such values of $p$ and $n$ allow the energy density of EDE dilutes faster than matter and quickly becomes irrelevant to the cosmic evolution, particularly when the Universe becomes matter-dominated. Moreover, the parameters in the potential need to be tuned to affect the evolution around radiation-matter equality when one tries to resolve the Hubble tension. Below we investigate what parameter values influence the evolution of the Universe, when and to what extent. To this end, we follow the evolution of the EDE field from the time just after the reheating has been completed, which is regarded as the initial time in our calculation. We note that, although we specify the energy scale of inflation, we do not need to assume an explicit form for  the inflaton potential in the following argument.

\begin{figure}[t]
    \begin{center}
        \includegraphics[width=7.5cm]{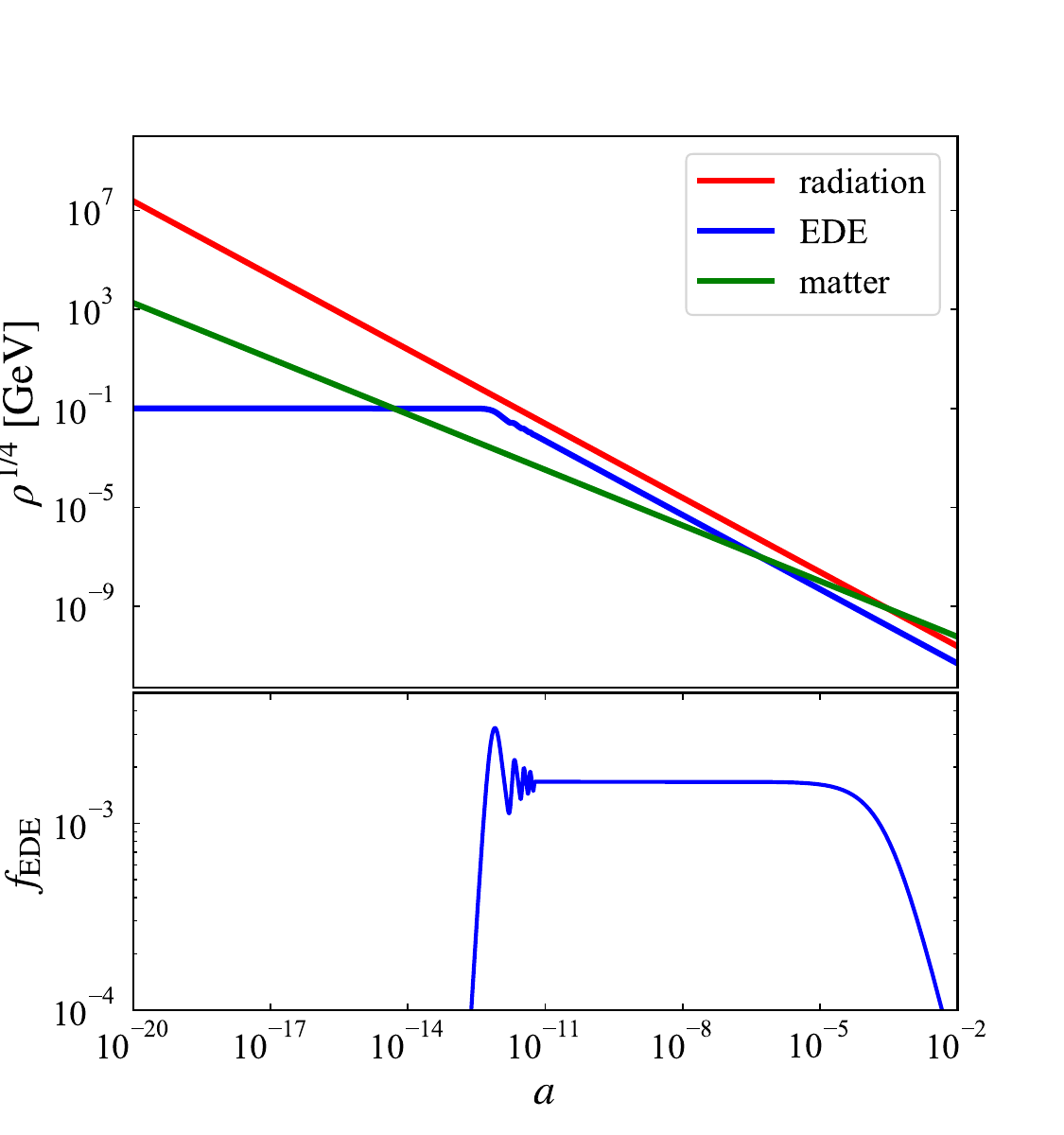}
        \includegraphics[width=7.5cm]{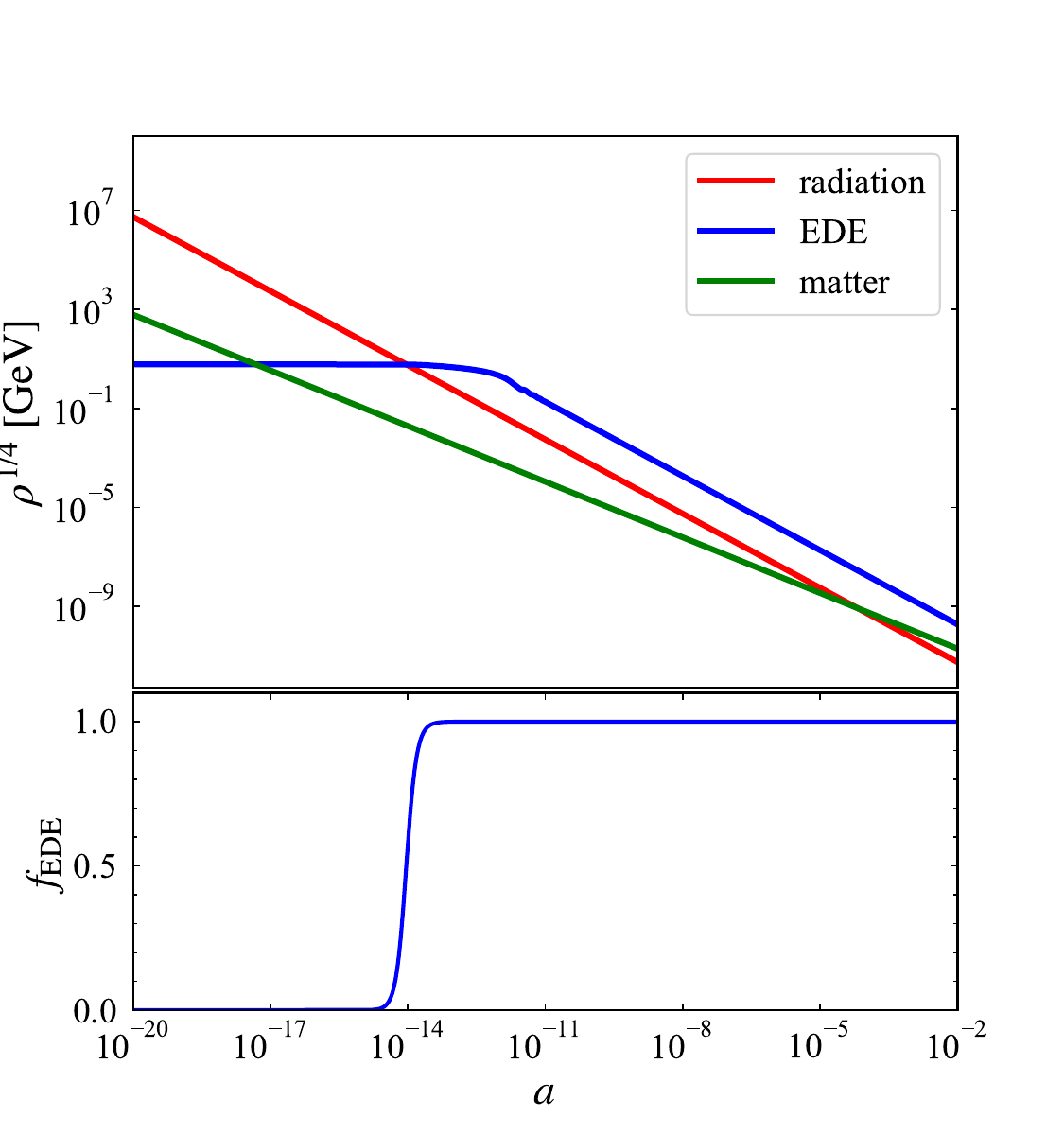}
        \caption{
        Evolution of energy densities of EDE, radiation and matter. In this figure, we assume the power-law potential with $p=4$ and take $\chi_{\rm ini}=0.1 M_{\rm Pl}$ and $V_0^{1/4}=1\,{\rm GeV}$ (left panel),  and $\chi_{\rm ini}=6 M_{\rm Pl}$ and $V_0^{1/4} = 1\,{\rm GeV}$ (right panel). 
        \label{fig:background1}} 
    \end{center}
\end{figure}
\begin{figure}[h]
    \begin{center}
        \includegraphics[width=7.5cm]{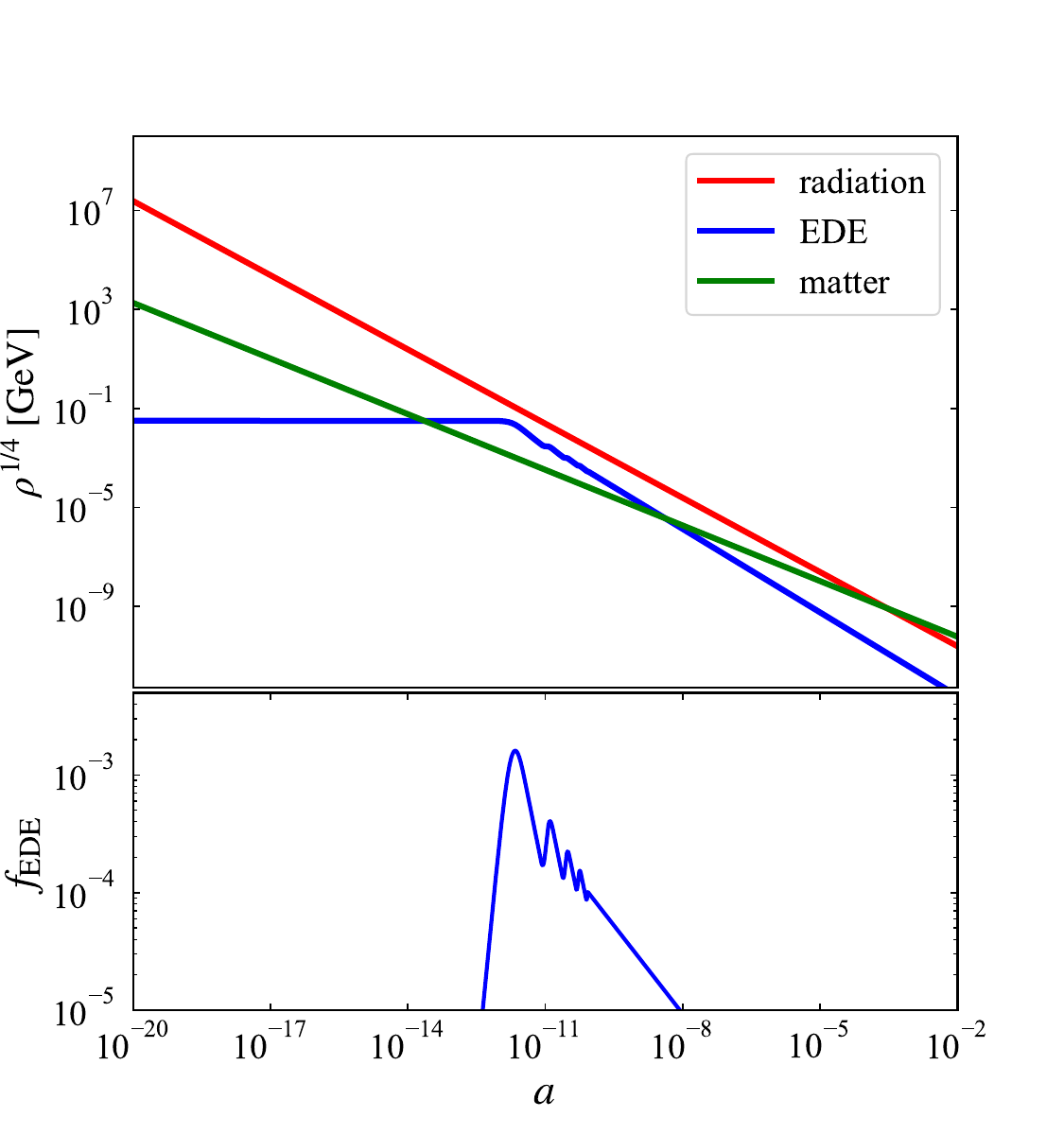}
        \includegraphics[width=7.5cm]{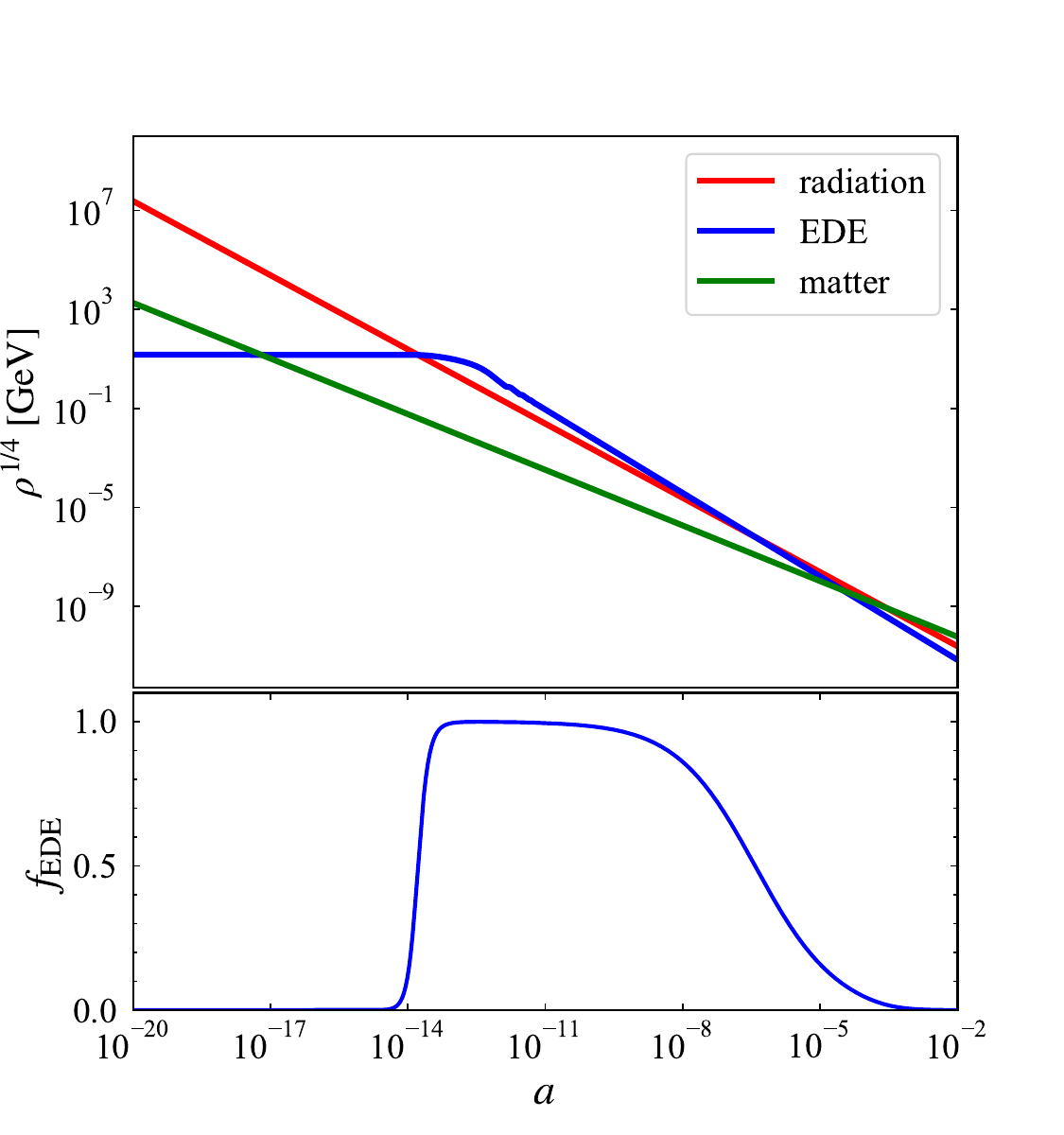}
        \caption{
        The same as Figure~\ref{fig:background1}, but for the power-law potential with $p=6$. Here we take $\chi_{\rm ini}=0.1 M_{\rm Pl}$ and $V_0^{1/4}=1\,{\rm GeV}$ (left panel) and $\chi_{\rm ini}=6 M_{\rm Pl}$ and $V_0^{1/4} = 1\,{\rm GeV}$ (right panel). 
        \label{fig:background2}} 
    \end{center}
\end{figure}
\begin{figure}[h]
    \begin{center}
        \includegraphics[width = 15cm]{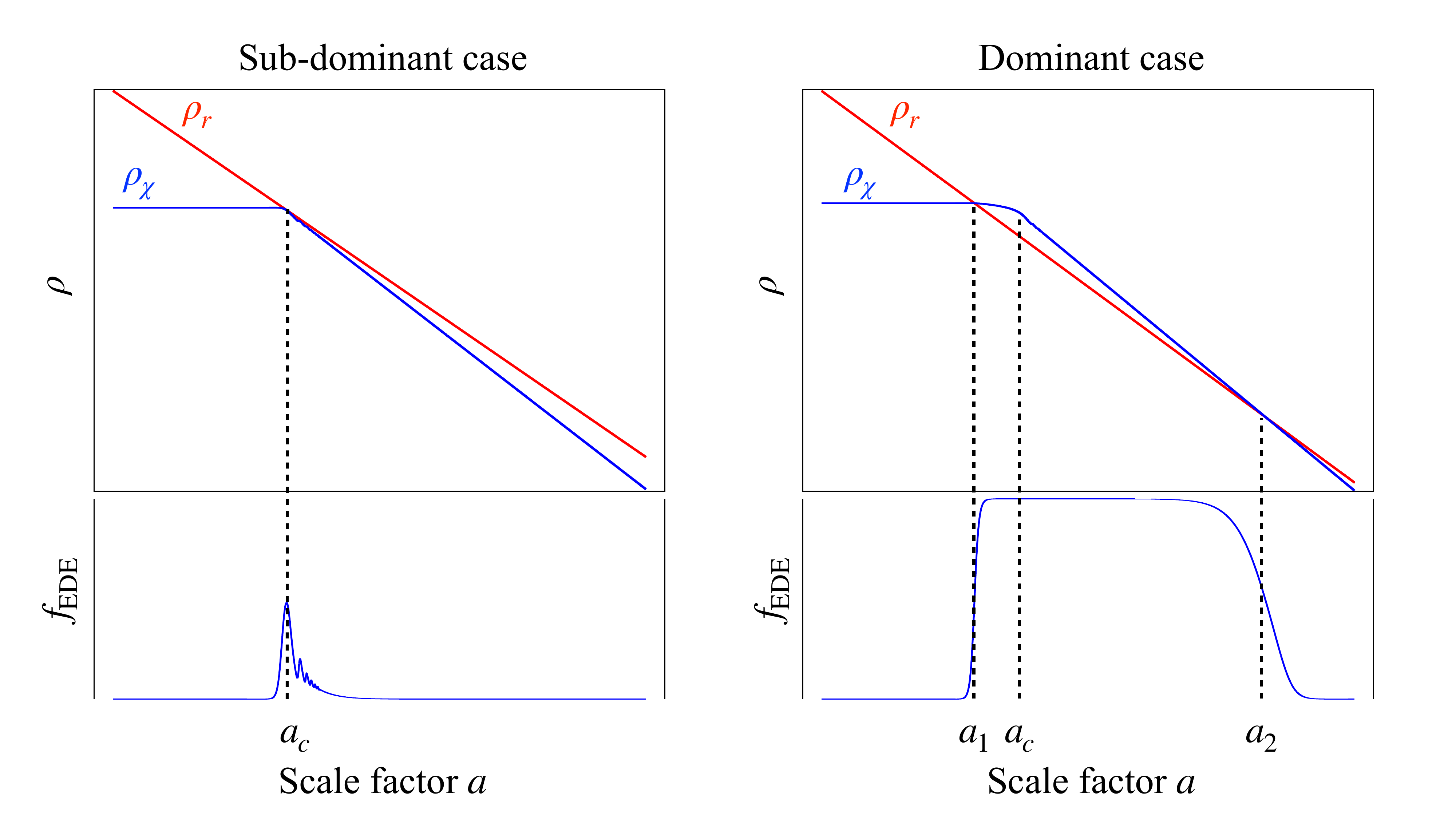}
        \caption{Schematic figure describing the characteristic scale factor $a_1, a_c$ and $a_2$ for the case of $p>4$. The right (left) panel corresponds to the case where the EDE field dominates at some epoch (always subdominant) during the course of the history of the Universe. 
        Red and blue lines describe energy densities of radiation and the EDE, respectively. The bottom panel shows the evolution of $f_{\rm EDE}$. The scale factor $a_c$ is defined as the one at which the $f_{\rm EDE}$ takes the maximum value. $a_1$ and $a_2$ correspond to the epoch at which the EDE energy density supersedes and is overtaken by that of radiation, respectively. Notice that $a_1$ and $a_2$ only appear when the EDE field dominates the Universe at some time (right panel). 
        \label{fig:Schematic}} 
    \end{center}
\end{figure}

In Figures~\ref{fig:background1} and \ref{fig:background2}, we show some examples of the thermal history in the generalized EDE model with the power-law potential. Since the energy density of an oscillating scalar field under the potential of $V(\chi) \propto \chi^{p}$ with $p>0$ scales as \cite{Shtanov:1994ce}
\begin{equation}
    \rho_\chi 
    \propto 
    a^{-6p / (p+2)} \,.
\end{equation}
Notice that $\rho_\chi$ for $p=4$ scales as the same as that of radiation, and when $p>4$, it dilutes faster than radiation. In the following argument, we also use the effective equation of state $w$ for the oscillating EDE field which is related to $p$ as 
\begin{equation}
    w 
    =
    \frac{p-2}{p+2} \,.    
\end{equation}
Here we only show the cases with the power-law potential since the axion-type potential~\eqref{eq:axion-type potential} around the minimum has a form $V(\chi) \propto \chi^{2n}$, and then the evolution of the axion-type EDE is quite similar to the one for the power-law type with $n=p/2$. In Figure~\ref{fig:background1}, the case for the power-law potential with $p=4$ is shown for $(\chi_{\rm ini},V_0^{1/4})=(0.1 M_{\rm Pl}, 1~{\rm GeV})$ (left panel) and $(6M_{\rm Pl}, 1~{\rm GeV})$ (right panel).  The case for the power-law potential with $p=6$ is also shown in Figure~\ref{fig:background2}, in which the values of $\chi_{\rm ini}$ and $V_0$ are taken to be the same as those in Figure~\ref{fig:background1}. In the left panel (in both Figures~\ref{fig:background1} and \ref{fig:background2}), we take the parameters such that the EDE field does not dominate the Universe during the whole history, on the other hand, the right panel corresponds to the case where the EDE field dominates the Universe during its slow-rolling phase and quasi-de Sitter phase appears before the EDE field starts to oscillate. Since $\rho_\chi$ for the case of $p=4$ decreases as the same as that of radiation, the Universe is dominated by EDE until matter does. It should be noted that a cosmological scenario where the oscillating EDE field dominates the Universe until matter supersedes it would be excluded by CMB observations, the case shown in the right panel of Figure \ref{fig:background1} is just for illustration purposes.

In the bottom panels of the figures, we also depict the evolution of $f_{\rm EDE}$ which represents the fraction of $\rho_\chi$ in the total energy density, defined as
\begin{align}\label{eq: fEDE def}
    f_{\rm EDE} \equiv \frac{\rho_\chi}{\rho_{\rm tot}}
    =
    \frac{\rho_\chi}{\rho_\chi + \rho_r + \rho_m + \rho_{\rm DE}}\,.
\end{align}
We also define the parameter $a_c$ and $\chi_c$ which denote the scale factor and the value of $\chi$ field at which $f_{\rm EDE}$ takes its maximum value\footnote{
This epoch roughly corresponds to the time when the EDE field starts to oscillate and $H \sim m_{\rm eff} \equiv \sqrt{V^{\prime\prime}}$ holds. However, this rough estimate fails especially when the quasi-de Sitter phase appears. Therefore we evaluate $a_c$ from the numerical calculation with the definition described here.
}. In Figure~\ref{fig:Schematic}, the schematic picture is shown to explain which epoch corresponds to $a_c$.

When the EDE dominates the Universe and quasi-de Sitter phase appears at some epoch as in the right panel of Figures~\ref{fig:background1} and \ref{fig:background2}, we define yet another scale factor (or time), denoted as $a_1$, at which the energy density of EDE supersedes that for radiation (see the right panel of Figure~\ref{fig:Schematic}).
The scale factor $a_1$ can be evaluated as 
\begin{align}
    \frac{a_1}{a_R} 
    \simeq 
    \bigg( 
    \frac{\rho_{r}(T_R)}{V(\chi_{\rm ini})} 
    \bigg)^{1/4}\,,
\end{align}
where $\chi_{\rm ini}$ is the initial value of $\chi$ field,  $a_R$ and $T_R$ are the scale factor and the temperature at the time of reheating, and $\rho_{r}(T_R)$ is radiation energy density at the reheating after inflation which is given by   
\begin{equation}
    \rho_r(T_R) 
    =
    \frac{\pi^2}{30} g_\ast (T_R) T_R^4 \,,
\end{equation}
with $g_\ast (T_R)$ the degrees of freedom at the time of reheating. In our numerical calculation, we assume that the inflationary Hubble scale is $H_{\rm inf}= 10^{13}\,{\rm GeV}$ and the reheating temperature is $T_R = 10^{15}\,{\rm GeV}$ for definiteness unless otherwise stated although their actual numbers do not affect our arguments.  The above choice of $H_{\rm inf}$ and $T_R$ almost corresponds to the case of the instantaneous reheating. For $p>4$, the energy density of an oscillating EDE field dilutes faster than radiation, and hence after the EDE field starts to oscillate, there appears the second equality when $\rho_\chi$ is overtaken by $\rho_r$, which we denote by $a_2$ (see the right panel of Figure~\ref{fig:Schematic}). We can express $a_2$ by using $a_c$ and the effective equation of state parameter $w$ for an oscillating EDE field as
\begin{align}
    \label{eq:a2}
    \frac{a_2}{a_c}
    \simeq 
    \Bigg[
    \bigg(\frac{a_c}{a_R}\bigg)^4
    \frac{V(\chi_{\rm c})}{\rho_r(T_R)}
    \Bigg]^{1/(3w - 1)}\,.
\end{align}
The analytic expression for $a_c$ is given in the next section.

\subsection{Estimates for $f_{\rm EDE}(a_c)$ and $a_c$ \label{subsec:fraction}}

In the context of the Hubble tension, $f_{\rm EDE}$ is an important parameter since the fraction of energy density of EDE determines its effects on the CMB power spectrum. Actually, as many analysis indicates, the EDE should give some fractional contribution to the total energy density of the Universe as $f_{\rm EDE} = {\cal O}(0.01)-(0.1)$ at around the radiation-matter equality, {\it i.e.}, $a_c \sim a_{\rm eq}$, to resolve the Hubble tension \cite{Poulin:2023lkg}.

\begin{figure}
    \centering
    \includegraphics[width=7.5cm]{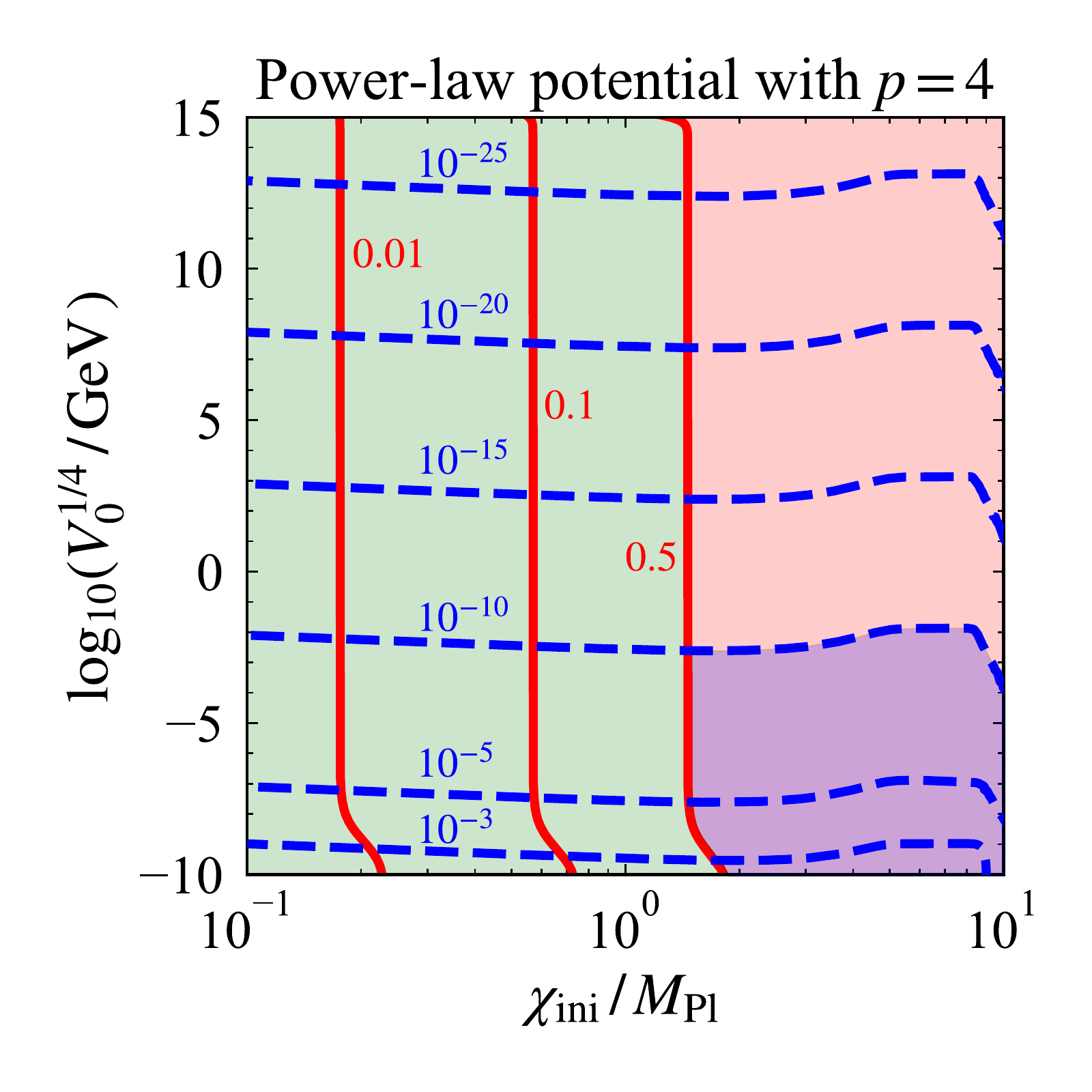}
    \includegraphics[width=7.5cm]{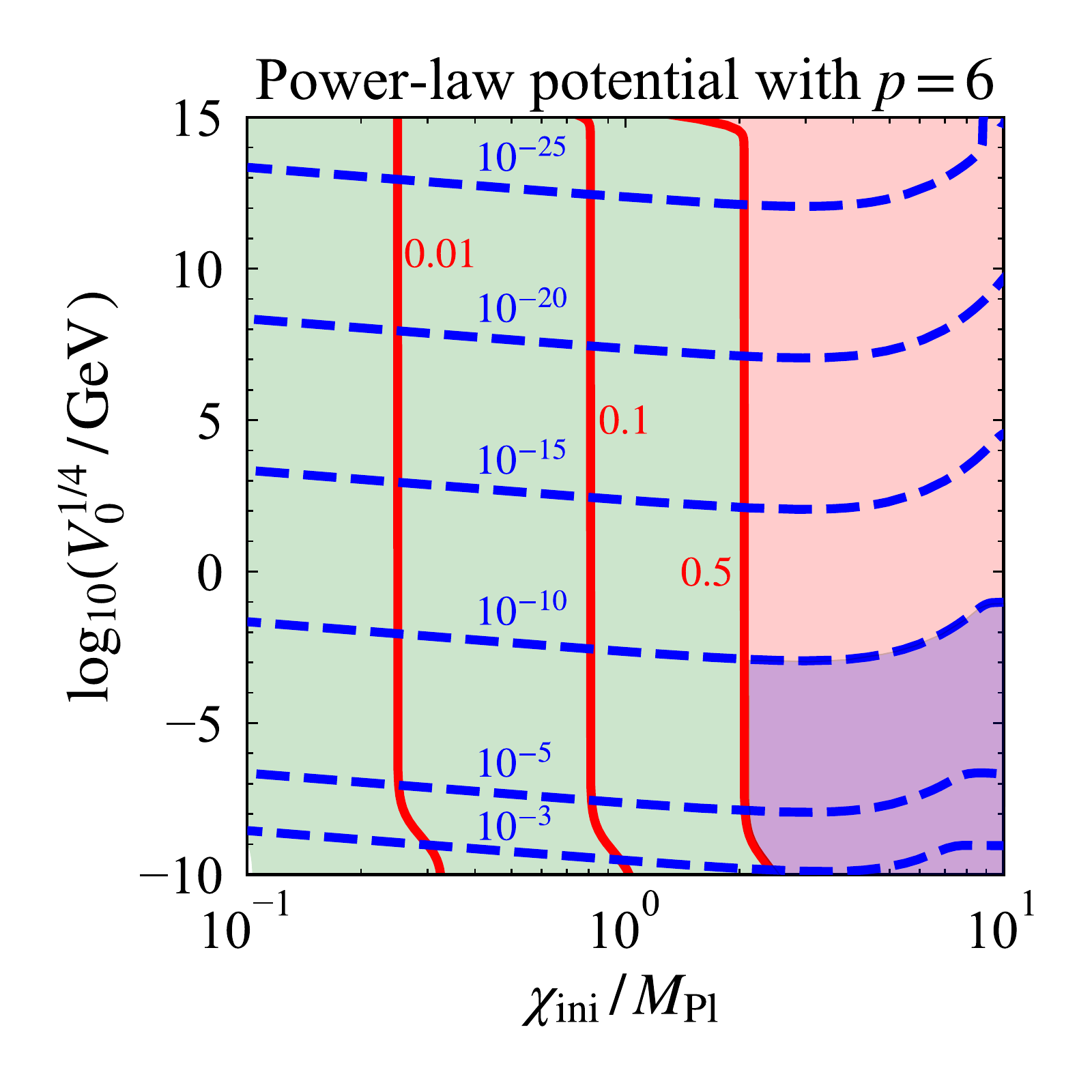}\\
    \includegraphics[width=7.5cm]{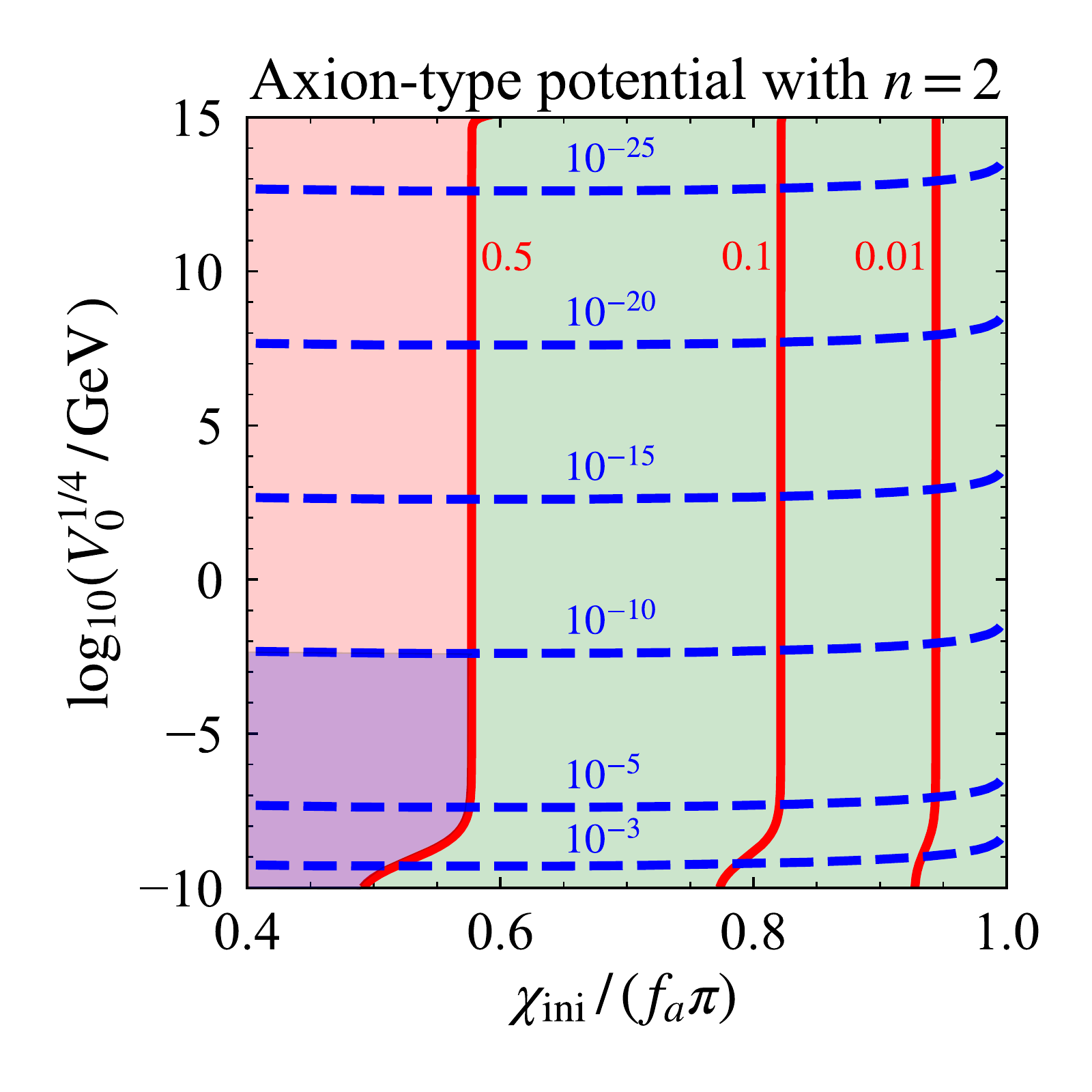}
    \includegraphics[width=7.5cm]{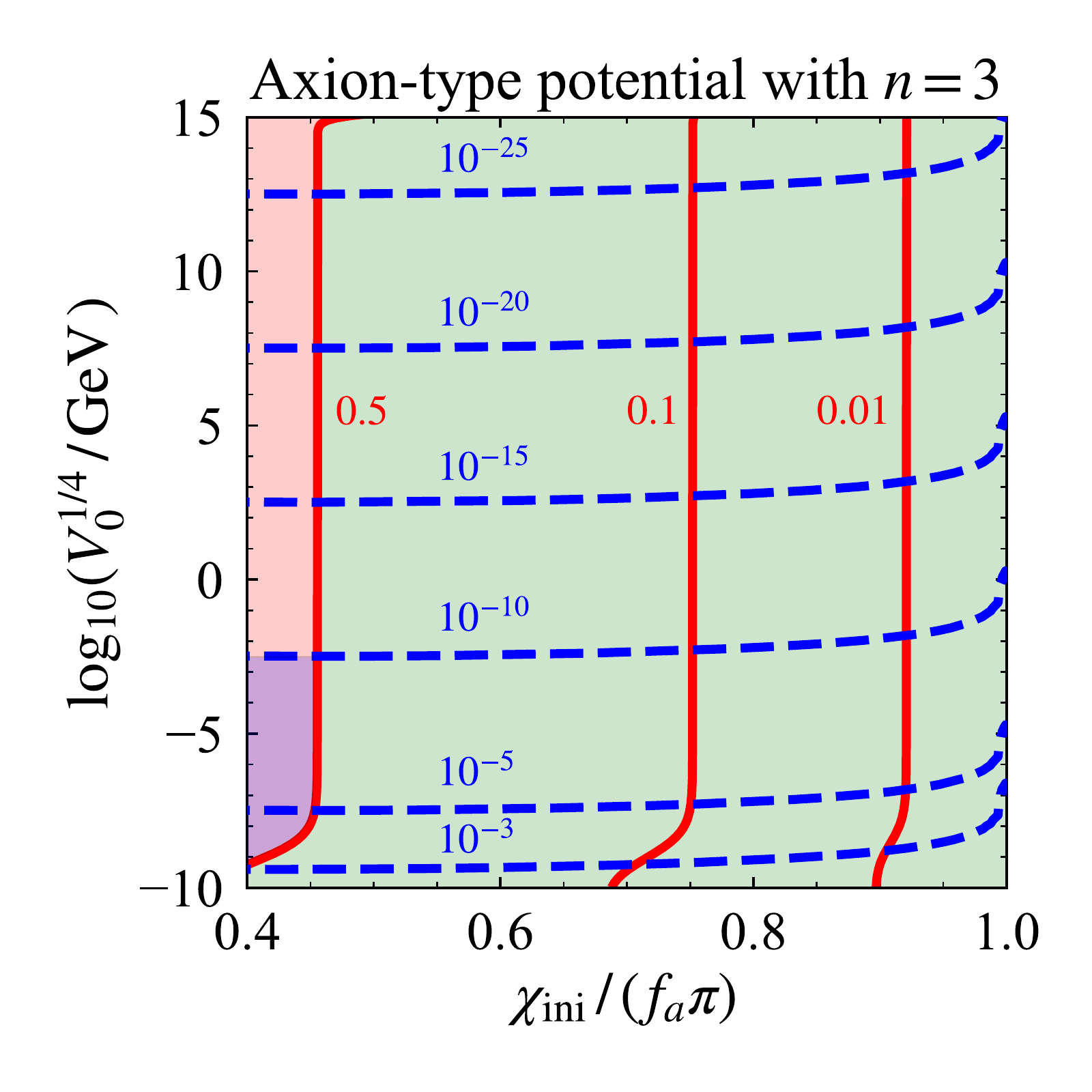}
    \caption{
    Contour plots of $f_{\rm EDE,c}$ (red) and $a_c$ (blue) in the $\chi_{\rm ini}$--$V_0$ plane. The upper and bottom panels show the power-law and the axion-type potentials with $p=2n=4$ (left panels) and $p=2n=6$ (right panels), respectively. The red and green region correspond to the case with $f_{\mrm{EDE},c}>0.5$ and $f_{\mrm{EDE},c} <0.5$, respectively. In the blue region, the EDE dominates after the BBN epoch, which would contradict cosmological observations. 
\label{fig:fcontour_fEDE}}
\end{figure}

Here we investigate what values of $f_{\rm EDE,c} (\equiv f_{\rm EDE} (a_c))$ and $a_c$ are obtained in a broad range of the parameter space. In Figure~\ref{fig:fcontour_fEDE}, we show contours of $f_{\rm EDE,c}$ and $a_c$ in the plane of $\chi_{\rm ini}$, and $V_0$ for the cases with the power-law (top panels) and the axion-type (bottom panels) potentials. When the EDE field dominates the Universe to generate the quasi-de Sitter phase after the epoch of BBN, the subsequent thermal history of the Universe is significantly changed, which would contradict cosmological observations. On the other hand, when such de Sitter phase appears before the BBN epoch, various cosmological constraints would be irrelevant and it can give an interesting implication for the GW observations which will be discussed in the next section. From the figure, one can also easily  see what values of $\chi_{\rm ini}$ and $V_0$ can realize $f_{\rm EDE,c} = {\cal O}(0.01)-(0.1)$ and $a_c = {\cal O}(10^{-4})$, which are necessary to resolve the Hubble tension.

Indeed one can analytically understand the behavior of $f_{\rm EDE,c}$ and $a_c$ in the $\chi_{\rm ini}$--$V_0$ plane as follows. First of all, $V(\chi_c)$ and $f_{\rm EDE,c}$ at $a=a_c$ are related as
\begin{align}
\label{eq:Vini_ac}
  V(\chi_{\rm c}) \simeq \frac{f_{\rm EDE,c}}{1 - f_{\rm EDE,c}}
  \rho_r(a_c)
  \,,
\end{align}
where we consider the case where the Universe is radiation-dominated at $a=a_c$, and we approximate the EDE energy density as $\rho_\chi (a_c) \simeq V (\chi_c)$. For the power-law potential\,\eqref{eq:power-law potential},  by taking the logarithm of both sides of Eq.~\eqref{eq:Vini_ac}, one obtains
\begin{align}
    \label{eq:power-law V0-chi relation}
    \log_{10} \left( 
    \frac{V_0^{1/4}}{M_{\rm Pl}} 
    \right) 
    + \frac{p}{4} \log_{10} \frac{\chi_c}{M_{\rm Pl}}
    = \frac{1}{4}\log_{10} \frac{f_{\rm EDE, c}}{1 - f_{\rm EDE, c}} + \log_{10} \left( 
    \frac{\rho_r(a_c)^{1/4}}{M_{\rm Pl}}
    \right) \,.
\end{align}
Actually $\rho_r (a_c)$ can be written with $\chi_{\rm ini}$ and $V_0$. Until the time when $a \simeq a_c$, the slow-roll approximation can be adopted for the equation of motion for the EDE field. By integrating Eq.~\eqref{eq: equation of motion chi} from  $a_R$ to $a_c$ under this approximation, one obtains
\begin{align}
    \label{eq:integrated_eom}
    - \int^{\chi_c}_{\chi_{\rm ini}} 
    \frac{\dd \chi}{V^\prime}
    \simeq 
    M_{\rm Pl}^2 \int^{a_c}_{a_R} 
    \frac{\dd\ln a}{\rho_{\rm tot}(a)}\,.
\end{align}
where the left-hand side of the above equation can be integrated as
\begin{align}
    -\int^{\chi_c}_{\chi_{\rm ini}} \frac{\dd \chi}{V^\prime(\chi)}
    \simeq
    \frac{1}{p(p - 2)}
    \frac{M_{\rm Pl}^2}{V_0} 
    \Bigg[
    \bigg(
    \frac{\chi_c}{M_{\rm Pl}}
    \bigg)^{-(p - 2)}
    -
     \bigg(
    \frac{\chi_{\rm ini}}{M_{\rm Pl}}
    \bigg)^{-(p - 2)}
    \Bigg]\,.
\end{align}
The evaluation of the right-hand side of \eqref{eq:integrated_eom} depends on whether the EDE dominates the Universe at $a=a_c$ or not, we discuss each case separately below.

\bigskip\bigskip
\noindent
{\bf $\bullet$ Case with EDE subdominant at $a_c$}
\vspace{1mm} \\ 
We first consider the case where the EDE is subdominant at $a=a_c$. In this case, the Universe is radiation-dominated between $a_R$ and $a_c$, and hence, by replacing $\rho_{\rm tot}$ with $\rho_r$ in the right-hand side of \eqref{eq:integrated_eom}, we obtain 
\begin{align}
\label{eq:EDE_subdominant_ac}
    \frac{1}{p(p - 2)}
    \frac{M_{\rm Pl}^2}{V_0} 
    \Bigg[
    \bigg(
    \frac{\chi_c}{M_{\rm Pl}}
    \bigg)^{-(p - 2)}
    -
    \bigg(
    \frac{\chi_{\rm ini}}{M_{\rm Pl}}
    \bigg)^{-(p - 2)}
    \Bigg]
    \simeq
    \frac{M_{\rm Pl}^2 }{4\rho_r(a_c)} \,,
\end{align}
where we have used the approximation that $\rho_r (a_R) \gg \rho_r(a_c)$. Putting the above expression into Eq.~\eqref{eq:power-law V0-chi relation}, we have
\begin{align}
    \label{eq: chiini RD}
    \frac{f_{\rm EDE, c}}{1 - f_{\rm EDE, c}}
    =
    \frac{4}{p(p - 2)} C^2 (1 - C^{p - 2})
    \bigg( \frac{\chi_{\rm ini}}{M_{\rm Pl}}\bigg)^2\,,
\end{align}
where we used $\chi_c = C \chi_{\rm ini}$ with $C$ being constant which holds for $\chi_{\rm ini} \leq \mathcal{O}(1)$, and our numerical analysis indicates that $C \simeq 0.65$. From Eq.\,\eqref{eq: chiini RD}, we can see that $f_{\rm EDE}$ does not depend on $V_0$ when the EDE is subdominant at $a=a_c$, {\it i.e.}, $f_{\rm EDE, c} < 0.5$ and $a_c < a_{\rm eq}$. From Eq.~\eqref{eq:EDE_subdominant_ac}, we can express $a_c$ by using $V_0$ and $\chi_{\rm ini}$ as 
\begin{align}\label{eq: ac analytical}
    \frac{a_c}{a_R}
    \simeq
    \Bigg[
    \frac{p(p - 2)}{4} \frac{V_0}{\rho_r(T_R)}
    \bigg(\frac{\chi_{\rm ini}}{M_{\rm Pl}}\bigg)^{p - 2}
    \Bigg]^{-1/4}\,.
\end{align}

We can also consider the case where $f_{\rm EDE}$ takes its maximum value during the matter-dominated epoch, namely $a_{\rm eq} < a_c$. In this case, by replacing $\rho_r (a_c)$ by $\rho_m (a_c)$ in Eq.\,\eqref{eq:Vini_ac} and $\rho_{\rm tot}(a)$ by $\rho_r(a) + \rho_m(a)$ in Eq.\,\eqref{eq:integrated_eom}, and then and integrating from $a_R$ to $a_c$, we can find that the $f_{\rm EDE,c}$ depends on both $V_0$ and $\chi_{\rm ini}$, contrary to Eq.~\eqref{eq: chiini RD} where $a = a_c$ occurs during radiation-dominated epoch. The dependence on $V_0$ and $\chi_{\rm ini}$ can be found in Figure~\ref{fig:fcontour_fEDE}.

\bigskip\bigskip
\noindent
{\bf $\bullet$ Case with EDE dominant at $a_c$}
\vspace{1mm} \\ 
In this case, we can integrate the right-hand side of \eqref{eq:integrated_eom}, for $a_c \ll a_{\rm eq}$, as 
\begin{align*}
    M_{\rm Pl}^2 \int_{a_R}^{a_c} \frac{\dd \ln a}{\rho_r(a) + \rho_\chi}
    &\simeq
    M_{\rm Pl}^2 \Bigg[
    \int_{a_R}^{a_1}\frac{\dd \ln a}{\rho_r(a)}
    + \int_{a_1}^{a_c}\frac{\dd \ln a}{\rho_\chi}
    \Bigg]
    \nonumber \\
    &\simeq
    M_{\rm Pl}^2\Bigg[\frac{1}{4\rho_r(a_1)} + \frac{1}{V(\chi_c)} \ln \frac{a_c}{a_1}\Bigg]\,.
\end{align*}
Thus, the scale factor $a_c$ is roughly estimated by
\begin{align}
    \frac{a_c}{a_1}
    \simeq
    \exp[\frac{p^{p-1}}{2^{p/2}(p -2)} \bigg(\frac{\chi_{\rm ini}}{M_{\rm Pl}}\bigg)^{-p} \bigg\{ \bigg(\frac{\chi_{\rm ini}}{M_{\rm Pl}}\bigg)^2 - \frac{p(p - 2)}{4} \bigg\} ]\,,
\end{align}
where we have expressed $\chi_c$ with $\chi_{\rm ini}$ by using the same procedure as done for the standard inflation case (see, {\it e.g.}, \cite{Lyth:2009zz}). The above expression can be inserted to Eq.\,\eqref{eq:a2} to obtain~$a_2$.

\bigskip\bigskip
From the above argument, one can see that $f_{\rm EDE,c}$ only depends on $\chi_{\rm ini}$ when $a_c < a_{\rm eq}$, which explains the behavior of the contours of $f_{\rm EDE,c}$ in most region of Figure~\ref{fig:fcontour_fEDE}. Although we have considered the power-law potential case in the above argument, the same also applies to the axion-type potential \eqref{eq:axion-type potential}, which explains the behavior of $f_{\rm EDE,c}$ in the bottom panels of Figure~\ref{fig:fcontour_fEDE}.

\subsection{Estimate of $\chi_{\rm ini}$ from stochastic argument}
\label{subsec:estimate of chi_ini}

The EDE field $\chi$ considered in this paper can be regarded as a spectator field whose contribution to the energy density is negligible during the inflationary era. When a spectator field is light enough, the quantum diffusion drives the distribution of its field value  to reach an equilibrium one, which can be discussed based on the stochastic formalism \cite{Starobinsky:1994bd,Starobinsky:1986fx,Enqvist:2012xn,Hardwick:2017fjo} and a typical value of $\chi_{\rm ini}$ can be inferred given the inflationary energy scale $H_{\rm inf}$ and the parameter in the potential of $\chi$. Here we briefly discuss such a typical value of $\chi_{\rm ini}$.

The field value of $\chi$ follows Langevin equation:
\begin{align}
    \dv{\chi(N)}{N}
    =
    - \frac{V_{,\chi}(\chi)}{3H^2} 
    + \frac{H}{2\pi}\,\xi(N)\,,
\end{align}
where we take the number of $e$-fold $N \equiv \ln a$ as a time variable and $\xi(N)$ is a Gaussian white noise. The first and second terms on the right-hand side correspond to classical motion and quantum fluctuations, respectively.

From the above equation, we can get the Fokker-Planck equation as \cite{Starobinsky:1986fx},
\begin{align}
    \pdv{P(N, \chi)}{N}
    =
    \pdv{\chi}
    \Bigg[
    \pdv{V(\chi)}{\chi} \frac{P(N, \chi)}{3H^2}
    + \frac{H^2}{8\pi^2} \pdv{P(N, \chi)}{\chi}
    \Bigg]\,,
\end{align}
where $P(N, \chi)$ is the probability distribution function (PDF) of the field value of a spectator field $\chi$. An equilibrium solution for $P(N,\chi)$ can be found as \cite{Enqvist:2012xn,Hardwick:2017fjo}, 
\begin{align}
    \label{eq: static PDF}
    P_{\rm stat}(\chi) 
    \propto 
    \exp\Bigg[ 
    -\frac{8\pi^2V(\chi)}{3H^4_{\rm inf}} 
    \Bigg]\,.
\end{align}
Here, we assume that the PDF relaxes to an equilibrium stationary solution by the end of inflation. We can obtain a typical value of the spectator field by setting the absolute value of the exponent approximately equal to unity.

\begin{figure}[t]
    \begin{center}
        \includegraphics[width=7.5cm]{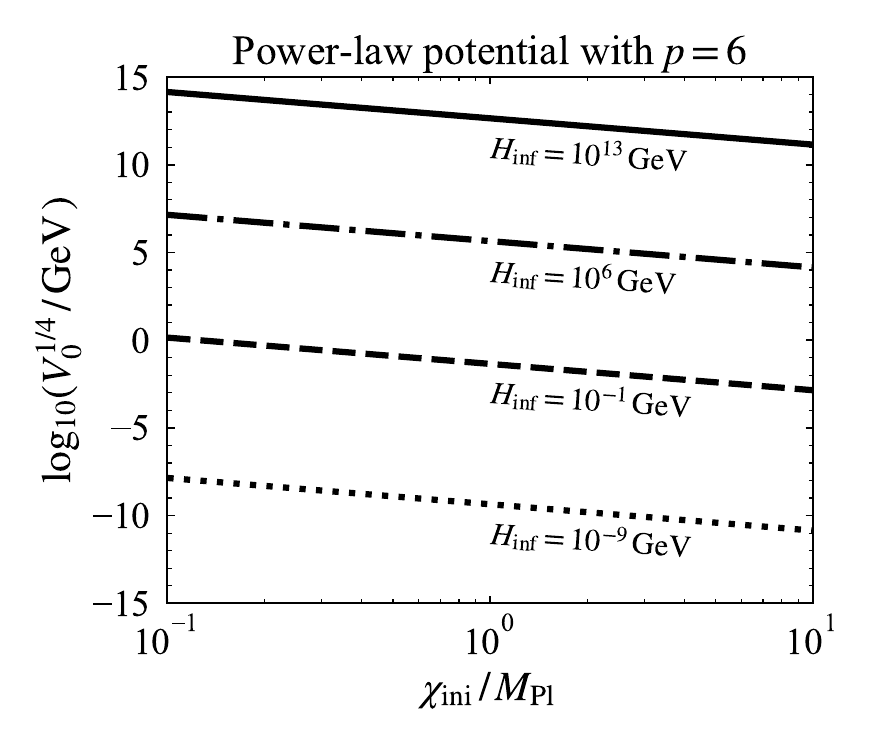}
        \includegraphics[width=7.5cm]{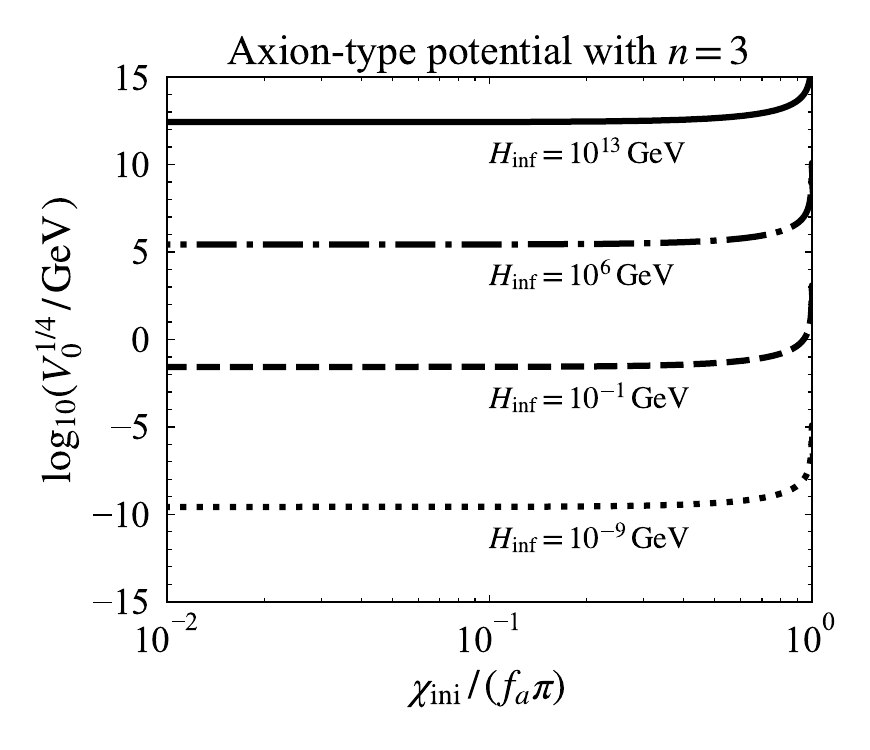}
        \caption{
        Contours of $H_{\rm inf}$ suggested from the stochastic argument in the $\chi_{\rm ini}$--$V_0$ plane for the power-law (left) and the axion-type (right) potentials with $p=2n=6$.
        \label{fig:stochastic}}
    \end{center}
\end{figure}

Based on the argument above, we can estimate the values of $\chi_{\rm ini}$ and $V_0$, which are depicted in Figure~\ref{fig:stochastic} for a given $H_{\rm inf}$. From the figure, one can see that when the inflationary Hubble scale is $H_{\rm inf}=10^{13}\,{\rm GeV} ~ (10^{6}\,{\rm GeV} )$, the scale factor at which the EDE takes its maximum contribution to the total energy density is $a_c = {\cal O}(10^{-26}) - {\cal O}(10^{-25})~ \bigl({\cal O}(10^{-19}) - {\cal O}(10^{-18})\bigr)$ for the power-law potential, which is much earlier than the CMB and even BBN epoch. As already mentioned, to resolve the Hubble tension in the framework of the EDE model, one needs to have $a_c = {\cal O} (10^{-4})$.  Figure~\ref{fig:fcontour_fEDE} indicates that this can be realized when $V_0^{1/4} \simeq 10^{-9}~{\rm GeV}$, which corresponds to a relatively low inflationary scale of $H_{\rm inf} = {\cal O}(10^{-9})\,{\rm GeV}$ as seen from Figure~\ref{fig:stochastic}.

From the discussion here, one can notice that a low-scale inflation is suggested to resolve the Hubble tension, based the stochastic argument when the equilibrium distribution is reached during inflation\footnote{
Whether the equilibrium distribution is realized or not depends on the potential of the inflaton \cite{Hardwick:2017fjo}.
}. For the axion-type potential, we can also draw almost the same conclusion.

\section{Gravitational waves spectrum}
\label{sec:GW}
In this section, we discuss the consequences of the generalized EDE to the GW spectrum, particularly for the case where the EDE dominates the Universe at some point and, then subsequently its energy density dilutes faster than radiation, {\it i.e.}, $n>4$ and $p>2$ for the power-law and axion-type potentials, respectively\footnote{
Actulally resonant amplification of EDE field fluctuations can give sizable GW background in these kind of potentials \cite{Kitajima:2023mxn}. The frequency range is somewhat different from the one discussed here, but such GW background could be another signature of the generalized EDE.
}.

\subsection{Gravitational waves spectrum}
\label{subsec:GW specturm}

First we briefly describe how the GW spectrum is calculated following the standard procedure. The equation of motion for tensor perturbation in the transverse-traceless gauge in the Fourier space $h^{\lambda}_{\mathbf{k}}$  for the polarization $\lambda = (+, \times)$ is written by
\begin{align}
	\ddot{h}^{\lambda}_{\mathbf{k}}
	+ 3H \dot{h}^{\lambda}_{\mathbf{k}}
	+ \frac{k^2}{a^2}\,  h^\lambda_{\mathbf{k}}
	= 0\,.
\end{align}
The GW spectrum, which is the energy density of GWs normalized by the critical energy density per logarithmic interval, is given by 
\begin{align}
    \Omega_{\rm GW}(k) 
    = \frac{1}{12} 
    \bigg( \frac{k}{a H} \bigg) 
    \mathcal{P}_T(k) T_T^2(k) \,.
\end{align}
Here $T_T(k)$ is the transfer function and $\mathcal{P}_T(k)$ is the primordial tensor power spectrum, which is assumed
to have the power-law form expressed by
\begin{align}
   \mathcal{P}_T(k) 
   =
   A_T \bigg(\frac{k}{k_\ast} \bigg)^{n_T}\,,
\end{align}
where $A_T$ is the amplitude of the primordial GWs at the pivot scale $k_\ast$ and $n_T$ is the tensor spectral index. Here we choose the pivot scale as $k_\ast = 0.05~{\rm Mpc}^{-1}$. The amplitude of the tensor power spectrum can be determined by the inflationary energy scale, namely $\mathcal{P}_T = (8/M_{\rm Pl}^2) \big(H_{\rm inf}/2\pi\big)^2$. To describe the size of primordial GW spectrum, we usually use the tensor-to-scalar ratio which is defined by
\begin{align}
    r 
    =
    \frac{A_T}{A_S}\,,
\end{align}
where $A_S$ is the amplitude of the scalar primordial spectrum at the pivot scale $k_\ast$. In the following calculation, we take $r=0.036$ which corresponds to the 2$\sigma$ upper bound given by {\it Planck observation 2018} \cite{Planck:2018vyg} and {\it BICEP/Keck Collaboration 2018} \cite{BICEP:2021xfz} for illustration. Since the amplitude of the scalar power spectrum is given as $A_S=2.1\times10^{-9}$ \cite{Planck:2018vyg}, the above value of $r$ gives the energy scale of inflation as  $H_\mathrm{inf}\approx10^{13}\,{\rm GeV}$. Once $r$ is given, for the single-field inflation models, the tensor spectral index $n_T$ can be determined from the so-called consistency relation $n_T=-r/8$.

We numerically solve the equation of motion for $h^{\lambda}_{\mathbf{k}}$ to obtain the transfer function in models with the generalized EDE. The transfer function depends on the background equation of state \cite{Kuroyanagi:2008ye,Nakayama:2008wy,Nakayama:2009ce,Kuroyanagi:2014nba}, and the behavior of the GW spectrum can be easily captured by noticing that $\Omega_{\rm GW}$ scales as
\begin{align}
    \label{eq: Omega k dependence}
    \Omega_{\rm GW}
    \propto
    k^{2(3w - 1)/(3w + 1)}
    \propto 
    f^{2(3w - 1)/(3w + 1)}
    \,,
\end{align}
where $w$ is the equation of state parameter of the dominant component during the time when the mode $k$ enters the horizon and $f$ is the frequency corresponding to the mode~$k$. Since the effective equation of state parameter for an oscillating scalar field is given by $w = (p-2)/(p+2)$ for a power-law potential $V(\chi) \propto \chi^p$, the GW spectrum is enhanced during when the oscillating EDE dominates the Universe when $p > 4$ ($n>2$ for the axion-type potential)\footnote{
Actually, when $f_{\rm EDE,c}\sim 0.5$, the motion of an EDE field can induce an oscillation in the Hubble parameter, which can make some peaks/dips in the GW spectrum \cite{Ye:2023xyr}. However, in the case of $f_{\rm EDE,c} > 0.5$, the enhancement discussed here hides such an effect.
}.  We summarized the scaling of 
the energy density of the oscillating EDE and the GW spectrum in  Table~\ref{table: n, w, logarho, logkomegaGW}. 
%
\begin{table}
    \caption{
   The effective equation of state parameter for an oscillating scalar field $w$, the indices $\beta$ and $\gamma$ for the scaling of the oscillating EDE energy density $\rho_{\rm EDE}\propto a^\beta$ and the GW spectrum $\Omega_{\rm GW} \propto k^\gamma$, respectively,  for $p$ (power-law potential) or $n$ (axion-type potential).   
   \label{table: n, w, logarho, logkomegaGW}}
    \centering
    \begin{tabular}{|c|rrrrrr|}
        \hline\rule{0cm}{0.5cm}   
        $n,~ p/2$   & $1$       & $2$           & $3$       & $4$       & ~~~~$\cdots$  & $\infty$~~~      \\
        \hline\rule{0cm}{0.5cm}   
        $w$         & 0         & $1/3$         & $1/2$     & $3/5$     & $\cdots$  & 1~~~~             \\[6pt]
        $\beta$     &~~~~$-3$   &~~~~~$-4$      &~~$-9/2$   & $-24/5$   & $\cdots$  &~~~~$-6$~~~~       \\[6pt]
        $\gamma$    & $-2$      & $0$           & $2/5$     & $4/7$     & $\cdots$  & $1$~~~~           \\[4pt]
        \hline
    \end{tabular}
\end{table}

\begin{figure}[t]
    \begin{center}
        \includegraphics[width=\linewidth]{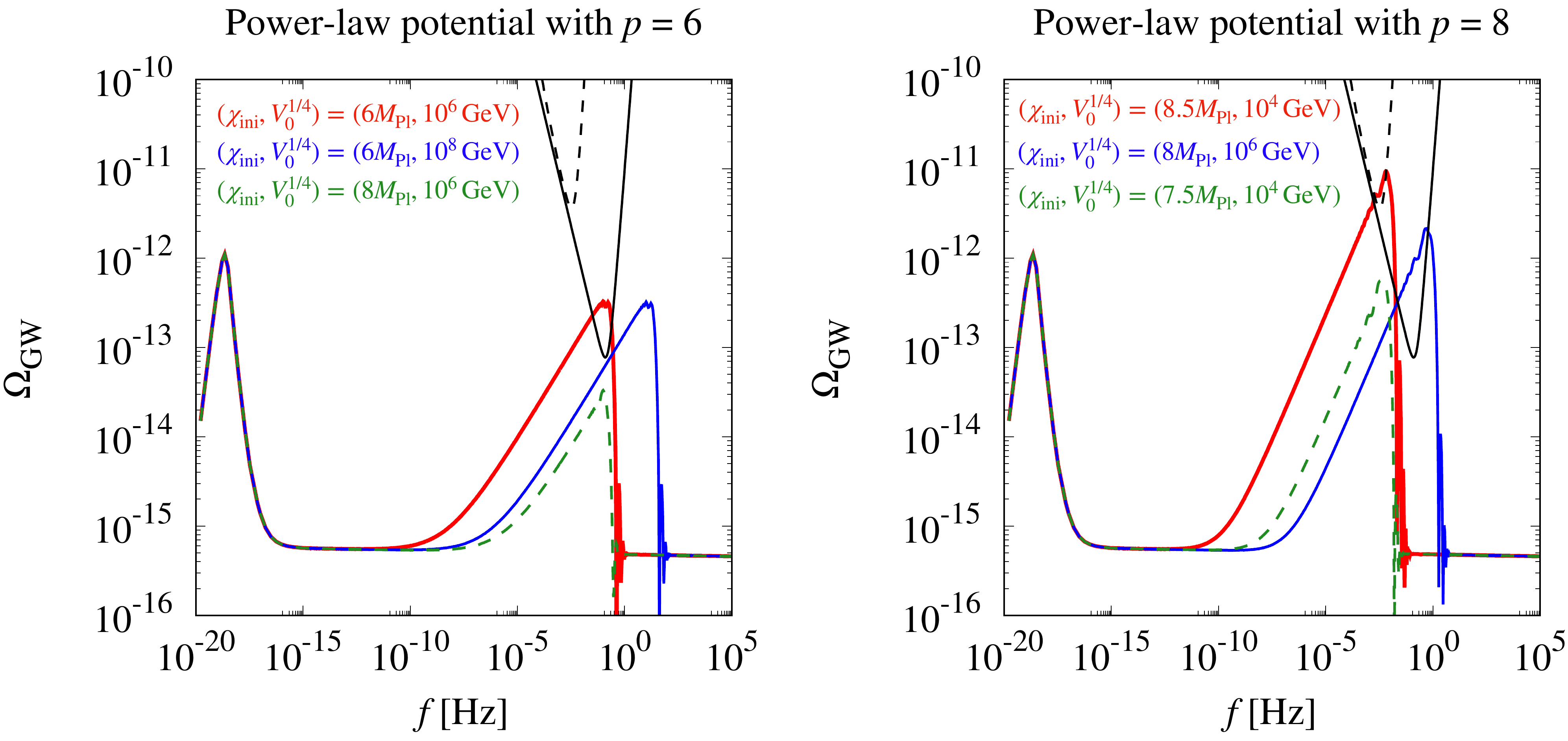}
        \caption{
        The GW spectrum for the case of the power-law potential with several values of $V_0$ and $\chi_{\rm ini}$ for the cases with $p=6$ (left) and $8$ (right). In the left panel, we take the parameters as $(\chi_{\rm ini}, V_0^{1/4}) = (6M_{\rm Pl}, 10^6\,{\rm GeV})$~(red), $(8M_{\rm Pl}, 10^8\,{\rm GeV})$~(blue), and $(8M_{\rm Pl}, 10^6\,{\rm GeV})$~(dashed green).  In the right panel, $(\chi_{\rm ini}, V_0^{1/4}) = (8.5 M_{\rm Pl}, 10^4~{\rm GeV})$~(red), $(8 M_{\rm Pl}, 10^6~{\rm GeV})$~(blue), and $(7.5 M_{\rm Pl}, 10^4~{\rm GeV})$~(dashed green). Black solid and dashed lines represent the sensitivity curves for DECIGO and LISA, respectively. 
        \label{fig:GWspectrum}} 
    \end{center}
\end{figure}

In Figure \ref{fig:GWspectrum}, we show the GW spectra for the power-law potential with several values of $\chi_{\rm ini}$ and $V_0$. We show the case with $p=6$ (left) and $p=8$ (right). For comparison, we also depict the sensitivity curves for future interferometer observations such as LISA \cite{Klein:2015hvg} and DECIGO \cite{Yagi:2011wg}. As seen from the figure, some range of the parameters can predict the GW signal detectable at LISA and DECIGO. Since the effective equation of state parameter $w$ for an oscillating EDE field is $w=1/2$ for $p = 6$ and $w = 3/5$ for $p = 8$, which indicates that the GW spectrum for the frequency mode which reenter the horizon during the oscillating EDE-dominated phase scales as $\Omega_{\rm GW} \propto k^{2/5}$ and $\Omega_{\rm GW} \propto k^{4/7}$ for $p=6$ and $8$, respectively as shown in Table~\ref{table: n, w, logarho, logkomegaGW}. A larger $p$ gives a steeper slope for the increase of the GW spectrum, and hence the case of a larger $p$ allows more parameter space for the detection of GWs. In particular, the case with $p=2n=\infty$ gives $\Omega_{\rm GW} \propto k$, which is the same as that for the kination one. GWs in models with kination have been studied in various context, for recent works, see {\it e.g.}, \cite{Co:2021lkc, Gouttenoire:2021jhk}.

Here we only show the GW spectrum for the power-law potential case, however, the case of the axion-type potential gives almost the same spectrum as that for the power-law one by identifying $n=p/2$. Therefore we do not show the axion-type potential case here.

\subsection{Detectable region in LISA and DECIGO}
\label{subsec:detectable region}

\begin{figure}[H]
    \begin{center}
        \includegraphics[width=12cm]{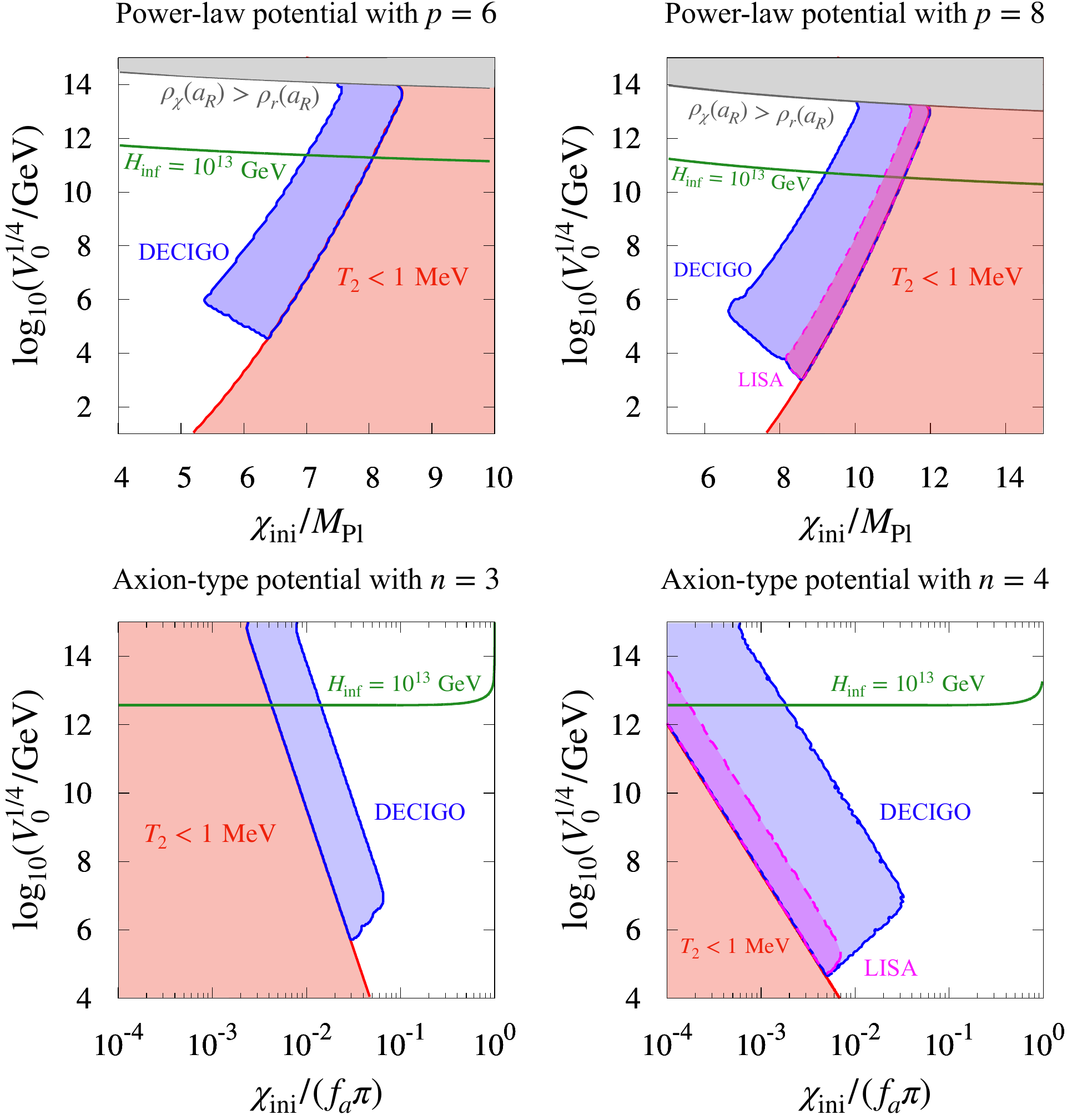}\\
        \caption{
        Detectable region in the $\chi_{\rm ini}$--$V_0$ plane. The above panels show the case of the power-law potential with $p = 6$ (top left) and $p = 8$ (top right), while the bottom ones show the case of the axion-type potential with $n = 3$ (bottom left) and $n = 4$ (bottom right). The magenta and blue regions correspond to the parameter space where GWs can be detected in LISA and DECIGO, respectively.
        The red region corresponds to $T_2 < 1\,{\rm MeV}$. The grey region does not satisfy our assumption that the generalized EDE field is subdominant at the time of reheating. The green line shows the prediction of $V_0$ and $\chi_{\rm ini}$ by the stochastic argument in section \ref{subsec:estimate of chi_ini}. 
        \label{fig:detectable}} 
    \end{center}
\end{figure}

In this section, we investigate what values of $\chi_{\rm ini}$ and $V_0$ can predict the stochastic GWs detectable in future observations such as LISA and DECIGO. As shown in Figure~\ref{fig:GWspectrum}, with some parameter choice, the GW spectrum in the generalized EDE model can be well above the sensitivity curves for LISA and DECIGO. In Figure~\ref{fig:detectable}, we depict the parameter region where the GW spectrum can be detected in LISA (magenta region) and DECIGO (blue region), {\it i.e.}, the spectrum can exceed the sensitivity curves of LISA and DECIGO, in the $\chi_{\rm ini}$--$V_0$ plane for the power-law potential with $p=6$ and 8 (top panels), and the axion-type potential with $n=3$ and 4 (bottom panels).  In the figure, we also show the parameter space where  $T_2 < T_{\rm BBN} \sim 1~{\rm MeV}$, in which the success of BBN would be spoiled since the Universe experiences a quasi-de Sitter phase during/after BBN in such a case. Furthermore, we only consider the case where the generalized EDE field is subdominant at the time of reheating, {\it i.e.,} $\rho_{\chi} (a_R) < \rho_r (a_R) $ . In the figure, grey region corresponds to the one where this condition is not satisfied. Since a larger $p$ ($n$) gives a steeper increase in the GW spectrum, the case of $p=8\, (n=4)$ shows more  parameter space for the detection of GWs in the power-law (axion-type) potential.

\clearpage
We also show the predicted values of $\chi_{\rm ini}$ and $V_0$ from the argument of the stochastic approach given in Section~\ref{subsec:estimate of chi_ini} for $H_{\rm inf}=10^{13}\,{\rm GeV}$. From the figure, we can see that when the equilibrium distribution for $\chi$ field is reached during inflation, a large (small) value of $\chi_{\rm ini}$ can give detectable GWs for the power-law (axion-type) potential with an appropriate choice of $V_0$.

\subsection{Implications for recent NANOGrav results
\label{subsec:nanograv}}

Recently the North American Nanohertz Observatory for Gravitational Waves (NANOGrav) reported evidence of the stochastic GW background from observations of the pulsar timing for 15 years\cite{NANOGrav:2023gor}, whose signal corresponds to $\Omega_{\rm GW}^{\rm (NANOGrav)} \approx 2.5\times10^{-8}$ at $f \approx 3.2\times10^{-8}~{\rm Hz}$. If the signal is generated from an inflationary stochastic GW background, one can interpret it with the primordial tensor power spectrum to be extremely blue-tilted, namely $n_T \simeq 1.8 \pm 0.3$ \cite{Vagnozzi:2023lwo}. Actually such a blue-tilted spectrum requires a low reheating temperature $T_R$ in order not to violate the BBN constraint \cite{Kuroyanagi:2020sfw}, and $T_R < 10~{\rm GeV}$ is demanded   \cite{Vagnozzi:2023lwo} from the recent NANOGrav signal. Indeed the existence of the generalized EDE can loosen the limit on $T_R$ and slightly lower $n_T$ to explain the NANOGrav signal. Here we briefly investigate to the implications of the generalized EDE for a blue-tilted spectrum provided that the inflationary GWs explain the NANOGrav signal.

\begin{figure}[t]
    \begin{center}
        \includegraphics[width=10cm]{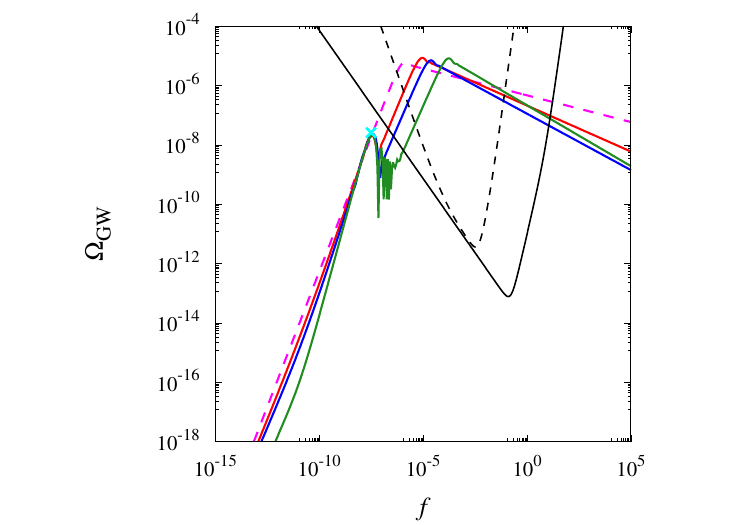}
        \caption{
        GW spectra for the power-law potential EDE with $p=6$ (red line), $p=8$ (blue line) and $p=\infty$ (green line)  for $r = 5.4\times 10^{-11}$. The values of $V_0, \chi_{\rm ini}$ and $n_T$ are assumed such that the spectra can explain the recent NANOGrav signal: $(V_0^{1/4}, \chi_{\rm ini}, n_T) = (10^{-0.7}\,{\rm GeV}, 4.3M_{\rm Pl}, 1.69)$ for $p=6$ and $ (10^{-1.25}\,{\rm GeV}, 6.2M_{\rm Pl}, 1.62)$ for $p=8$. For the case of $p=\infty$, we include such a generalized EDE as a fluid described in the text for the sake of numerical calculation. For comparison, the case without EDE with $n_T = 1.82$ (magenta dashed) is also shown. The cyan point represents  the recent NANOGrav 15-year signal.
        \label{fig:NANOGrav_VS_EDE_close_to_nT=1.8}} 
    \end{center}
\end{figure}

In Figure~\ref{fig:NANOGrav_VS_EDE_close_to_nT=1.8}, we show the GW spectra for the cases with and without the generalized EDE, both of which are assumed to have a blue-tilted tensor spectral index. In every case, the value of $n_T$ is taken such that the scale dependence of the GW spectrum at around NANOGrav frequency is $\Omega_{\rm GW} \propto f^2$ \cite{NANOGrav:2023gor,Vagnozzi:2023lwo}. We also assume the tensor-to-scalar ratio to be $r \simeq 5 \times 10^{-11}$ as in \cite{Vagnozzi:2023lwo}, and consider the power-law potential with $p = 6$ (red) and $p = 8$ (blue). Here we take the parameters as $(V_0^{1/4}, \chi_{\rm ini}, n_T) = (10^{-0.7}\,{\rm GeV}, 4.3 M_{\rm Pl}, 1.69)$ for $p=6$ and $(10^{-1.25}\,{\rm GeV}, 6.2 M_{\rm Pl}, 1.62)$ for $p=8$, respectively. These parameter sets satisfy the requirement that $T_2 > T_{\rm BBN} \simeq 1~{\rm MeV}$ to avoid the quasi-de Sitter phase after BBN. We also depict the GW spectrum for the case corresponding to $p \to \infty$ in which the scaling of the energy density during its oscillating phase is the same as that for the kination-dominated case. For the sake of numerical calculation, we include the case with $p \to \infty$ by adding an energy component which behaves as 
\begin{align}
    \rho_{\rm kin}(a)
    =
    \left\{
    \begin{aligned}
        & C_{\rm kin}  
        & \qquad (a \le a_c)  \,,\\        
        & C_{\rm kin} \bigg(
        \frac{a_c}{a}
        \bigg)^6  
        & \qquad (a>a_c) \,,
    \end{aligned}
    \right.
\end{align}
where $C_{\rm kin}$ is a constant and $a_c$ is the scale factor at which the kination phase starts. Here we take $C_{\rm kin} = 10^{1.6}~{\rm GeV}^4$, $a_c=10^{-12}$ and $n_T = 1.59$. For the argument involving the kination epoch to explain the NANOGrav results in different frameworks, see {\it e.g.}, \cite{Chowdhury:2023opo, Harigaya:2023pmw}.

It should be noticed that, with the existence of generalized EDE, the GW spectrum can be enhanced to on top of the blue-tilted primordial GWs. The GW spectrum without the EDE, which is  depicted by the dashed magenta line in Figure~\ref{fig:NANOGrav_VS_EDE_close_to_nT=1.8}, assumes $n_T \simeq 1.82$. On the other hand, in the generalized EDE model, the value of $n_T$ can be slightly lowered as shown in Figure~\ref{fig:NANOGrav_VS_EDE_close_to_nT=1.8}. Moreover, the bound on the reheating temperature is relaxed as $T_R <  150\,{\rm GeV}\,, 400\,{\rm GeV}$ and $5\times 10^3,{\rm GeV}$ for the cases with $p = 6\,,  8$ and $\infty$, respectively. This comes from the fact that the GW amplitude of the modes which enter the horizon during the quasi-de Sitter phase is abruptly suppressed, as shown in Figure~\ref{fig:GWspectrum}. This is a unique feature of the generalized EDE scenario and does not occur in models with a simple kination phase.

\section{Conclusion}
\label{sec:conclusion}

We have studied the consequences of a scalar field whose energy density can give a non-negligible contribution at some point during the course of the history of the Universe. Such kind of scalar field has recently been discussed as a potential solution to the Hubble tension and is called early dark energy (EDE). In such a EDE model, its energy density gives some contribution at around recombination, and then it quickly dilutes to become a negligible component in later time to be consistent with cosmological observations such as CMB. To realize this situation, the parameters in the scalar-field potential need to be fine-tuned. However, on general grounds, one can take a broad parameter range unless such a parameter choice is motivated by some arguments, and furthermore other cosmological aspects can be affected. We dubbed such an EDE field which can take broad parameter range as ``generalized EDE."

In this paper, two different types of potentials for the scalar field $\chi$ have been considered: power-law and axion-type ones, in which there are three free parameters: the power index $p$ (or $n$), the energy scale $V_0$, and the initial value $\chi_{\rm ini}$ of the scalar field.  We have assumed that the energy density of the scalar field is less than that of radiation at the time of reheating.

First, we have investigated to what extent and when the energy density of the generalized EDE field can be sizable as a function of $\chi_{\rm ini}$ and $V_0$, which was shown in Figure~\ref{fig:fcontour_fEDE}. As discussed in Section~\ref{sec:background}, the EDE field can give a non-negligible contribution to the total energy density in a broad parameter range, particularly when $\chi_{\rm ini} > {\cal O}(0.1)\,M_{\rm pl}$ $( \chi_{\rm ini} < {\cal O}(0.1)\,\pi f_a)$ for the power-law (axion-type) potential. Then we studied what values of $\chi_{\rm ini}$ and $V_0$ are suggested from the argument of the stochastic formalism given the inflationary energy scale $H_{\rm inf}$. In order that the EDE can act as a possible solution to the Hubble tension, one needs $f_{\rm EDE,c} = {\cal O} (0.01)$ and $a_c = {\cal O}(10^{-4})$, which can be realized when $\chi_{\rm ini}=\mcal{O}(0.1)\,M_{\rm Pl}$ and $V_0^{1/4}\sim 10^{-9}\,{\rm GeV}$ for the power-law potential and $\chi_{\rm ini}/f_a\sim 0.9\pi$ and $V_0^{1/4}\sim 10^{-8}\,{\rm GeV}$ for the axion-type potential. To realize these values, from Figure~\ref{fig:stochastic}, one can see that a low inflationary scale as $H_{\rm inf} = {\cal O}(10^{-9})\,{\rm GeV}$ is suggested from the stochastic formalism argument.

We have also investigated the spectrum of the GW  background with the existence of the generalized EDE. We have shown that the GW spectrum is amplified as seen in Figure~\ref{fig:GWspectrum} if the EDE becomes dominant at some epoch, {\it i.e.}, $f_{\rm EDE}>0.5$. The enhancement of the spectrum almost depends on the initial value $\chi_{\rm ini}$ which controls the duration of the quasi-de Sitter phase. We studied the parameter ranges for $\chi_{\rm ini}$ and $V_0$ where the GW can be detected by future observations such as LISA and DECIGO, which is shown in Figure~\ref{fig:detectable}.

Finally, we have briefly discussed the implications of the generalized EDE for the NANOGrav 15-year signal, which indicates that $\Omega_{\rm GW}^{\rm (NANOGrav)}\approx 2.5\times 10^{-8}$ at  $f \approx 3.2\times10^{-8}~{\rm Hz}$. Assuming that the inflationary GWs can explain the signal, one needs a very blue-tilted primordial tensor power spectrum. In the standard case ({\it i.e.}, without the generalized EDE),  the tensor spectral index $n_T$ should be as large as $n_T \simeq 1.8$ for $r \simeq 5\times 10^{-11}$ to be well fitted to the signal as in \cite{Vagnozzi:2023lwo}. It should also be noted that, with such a blue-tilted spectrum, the reheating temperature needs to be lowered not to contradict with the BBN constraint and $T_R < 10\,{\rm GeV}$ is required \cite{Vagnozzi:2023lwo}. However, with the existence of the EDE and appropriate parameter choices, we found that $n_T$ can be reduced to $n_T = 1.69$ for $p = 6$, $n_T = 1.62$ for $p = 8$, and $n_T = 1.59$ for $p=\infty$. Besides, we also found that the EDE cane relax the bound on the reheating temperature to $T_R = 150\,{\rm GeV}$ for $p = 6$, $T_R = 400\, {\rm GeV}$ for $p = 8$, and $T_R = 1.59$ for $p=\infty$, which can be compared to the case of the standard thermal history $T_R < 10\,{\rm GeV}$ \cite{Vagnozzi:2023lwo}.

Scalar fields are predicted to ubiquitously exist in the early Universe in the light of high energy theories. The results of this work would help to consider the effects of such a scalar field on the evolution of the Universe.

\section*{Acknowledgment}
This work was supported by JSPS KAKENHI Grant Number 19K03874~(TT), 23K17691~(TT) and MEXT KAKENHI 23H04515 (TT). 

\bibliography{GW_EDE}
\end{document}